\documentclass[aps,prx,twocolumn,superscriptaddress,showpacs,floatfix]{revtex4}

\usepackage{amsfonts}

\usepackage{subfigure}

\usepackage{graphicx}
\usepackage{dcolumn}
\usepackage{bm}
\usepackage{hyperref}
\usepackage[mathlines]{lineno}
\usepackage{amsmath}

\preprint{}

\begin{document}

\title{Natural Orbitals Renormalization Group Approach to
the Two-Impurity Kondo Critical Point}

\preprint{1}

\author{Rong-Qiang He}
\affiliation{Department of Physics, Renmin University of China, Beijing 100872, China}
\affiliation{Institute for Advanced Study, Tsinghua University, Beijing 100084, China}
\author{Jianhui Dai}
\affiliation{Department of Physics, Hangzhou Normal University, Hangzhou 310036, China}
\author{Zhong-Yi Lu}
\affiliation{Department of Physics, Renmin University of China, Beijing 100872, China}

\date{\today}

\begin{abstract}

The problem of two magnetic impurities in a normal metal exposes the two opposite tendencies in the formation of a singlet ground state, driven respectively by the single-ion Kondo effect with conduction electrons to screen impurity spins or the Ruderman-Kittel-Kasuya-Yosida interaction between the two impurities to directly form impurity spin singlet. However, whether the competition between these two tendencies can lead to a quantum critical point has been debated over more than two decades. Here, we study this problem by applying the newly proposed natural orbitals renormalization group method to a lattice version of the two-impurity Kondo model with a direct exchange $K$ between the two impurity spins. The method allows for unbiased accessing the ground state wave functions and low-lying excitations for sufficiently large system sizes. We demonstrate the existence of a quantum critical point, characterized by the power-law divergence of impurity staggered susceptibility with critical exponent $\gamma = 0.60(1)$, on the antiferromagnetic side of $K$ when the interimpurity distance $R$ is even lattice spacing, while a crossover behavior is recovered when $R$ is odd lattice spacing. These results have ultimately resolved the long-standing discrepancy between the numerical renormalization group and quantum Monte Carlo studies, confirming a link of this two-impurity Kondo critical point to a hidden particle-hole symmetry predicted by the local Fermi liquid theory.

\end{abstract}

\pacs{75.10.Hf,71.15.Dx,75.20.Hr,75.40.Cx}

\maketitle

\section{Introduction}

It is well-recognized that the competition between the single-ion Kondo effect and the Ruderman-Kittel-Kasuya-Yosida (RKKY) interaction, the two inevitable forces in any Kondo systems with more than one local impurity magnetic moment, plays a crucial role in correlated systems ranging from dilute magnetic alloys to heavy fermion compounds \cite{Hewson93,Coleman07}. The indirect RKKY interaction, namely the interimpurity interaction mediated by conduction electrons via a short-range (on-site) Kondo coupling, oscillates and decays with the interimpurity distance $R$ and Fermi momentum $2k_F$ \cite{RK54,Kasuya56,Yosida57}. When the RKKY interaction grows toward the strong antiferromagnetic limit, the quantum many-body ground state will evolve from a collective Kondo singlet state\cite{Yosida66} into the interimpurity singlet state locked by the RKKY interaction. However, whether a distinct separation or a quantum critical point exists between the two singlet ground states has remained elusive. This theoretical issue has a fundamental importance as it closely correlates with the critical divergence or scaling behavior of several physical quantities, or emergent energy scales in realistic materials\cite{Si08,Yang08,Zhu11}.

It is remarkable that even for the simplest case with only two local impurities, i.e., the two-impurity Kondo model (TIKM)\cite{JKW81}, the evidences for a quantum critical point separating the two distinct singlet states do not converge. On the one hand, the early
numerical renormalization group (NRG) studies\cite{Jones87,Jones88,Jones89a} revealed an unstable fixed point characterized by diverging impurity staggered susceptibility and coefficient of specific heat at a finite ratio of the RKKY interaction to the Kondo temperature ($\approx 2.2$). On the other hand, the subsequent well-controlled quantum Monte Carlo (QMC)
studies did not find such a divergence, instead, a crossover behavior in the corresponding quantities at very low but still finite temperatures was observed \cite{Fye89,Fye94}. Since then, although the effective field theory analysis suggested the occurrence of critical points in several variants of the TIKM \cite{Jones89b,Affleck92,Sire93,Gan95,Zarand06,note1}, the strong debate in the TIKM studies, especially in numerical studies, has still remained until today \cite{Sakai90,Sakai92,Silva96,Allerdt14}.


Almost all the previous investigations, except for the QMC studies \cite{Fye89,Fye94}, rely on a decomposition of the conduction electrons into odd/even channels with respect to the impurity center. Under such a decomposition, the original two-impurity problem is mapped effectively onto two-channel/two-impurity problem defined on the momentum or energy space, resulting in various impurity couplings which are energy-dependent in general. The obtained two-channel/two-impurity model, capturing the low-energy properties of the original problem, provides a base for the NRG and effective field theory studies. In particular, the early NRG
calculations \cite{Jones87,Jones88} assume ``energy-independent'' coupling constants in odd/even channels. The QMC simulations \cite{Fye89,Fye94}, on the other hand, suffer from finite temperatures. Other NRG\cite{Sakai90,Sakai92,Silva96} or DMRG\cite{Allerdt14} studies, where no evidence was found for a quantum critical point after taking into account the
``energy-dependent'' coupling strengths or starting from a real-space two-impurity lattice model, indeed show the importance of lattice geometry details.

More seriously, these numerical results seem to strongly contradict a general phase shift argument based on the local Fermi liquid theory\cite{Nozieres74,Nozieres80}, which states that a phase transition must exist between the two stable fixed points if a TIKM preserves particle-hole (PH) symmetry \cite{Millis90,Affleck95}. As a matter of fact, there are two types of PH symmetries associated with a standard TIKM, corresponding to the cases with the interimpurity distance $R$ being even or odd, or in the lattice case, the impurity mirror center being on site or bond, respectively. It is the first type of PH symmetry, namely the
distance $R$ being even or the mirror center being on site, that can guarantee a phase transition \cite{Affleck95,suppl}. Although the phase shift argument comes from the odd/even decomposition with a number of simplifications including the spherical plane wave and linear dispersion approximations, its validity should not depend on ignoring energy dependence of coupling constants and other lattice details. Otherwise, if there is no any phase transition in this case, it would indicate a breakdown of the local Fermi liquid picture of the single-impurity Kondo problem. Here we would like to emphasize that such a phase transition could be either a first-order transition or a quantum critical point. In the case of only the second type of PH symmetry preserved, there is no such a guarantee, in other words, either a phase transition or a crossover takes place between the two stable fixed points.

In our point of view, the most straightforward approach to challenge or confirm this argument and further clarify the discrepancies among various numerical studies is to directly solve the ground state of a standard TKIM for sufficiently large systems without using decomposition. In fact, Affleck et al.\cite{Affleck95} already outlined three conditions for such a decisive numerical study: (1) the studied model preserves the first type of PH symmetry; (2) a model parameter is varied to pass the interimpurity singlet to the Kondo singlet; (3) sufficiently low temperatures are accessible and other model parameters are fine-tuned.

In this paper, we reexamine whether or not there is a quantum critical point in a standard TIKM by applying a newly developed numerical method, i.e., the natural orbitals renormalization group (NORG)\cite{He14}. Different from the conventional NRG, the NORG keeps faithfully all the lattice geometry details and does not need a decomposition into odd/even channels and mapping onto the momentum or energy space. In particular, the ground state wave function as well as the relevant low energy excitations in electronic bath with several impurities can be calculated accurately in a well-controllable manner for a very large system. Therefore, the NORG is particularly pertinent for studying the zero temperature quantum phase transition in the two-impurity Kondo problem.

In the following we will present the numerical results and discuss the main observations. A brief description of the NORG method and an illustration of the phase shift argument and other relevant numerical results are given in the appendixes.

\begin{figure}
\includegraphics[width=8.0cm]{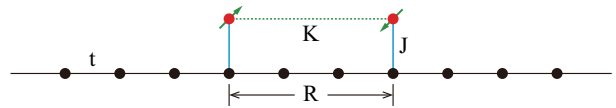}
\caption{(Color online) Schematic view of a standard two-impurity Kondo model with impurity separation $R = 3$.}
\label{model}
\end{figure}

\section{Model and Results}

Our studied TIKM is a standard one-dimensional lattice model\cite{note0}, as schematically shown in Fig. \ref{model}, which is described by the following Hamiltonian,
\begin{equation}\label{eq-hmlt}
\begin{array}{l}
H = {H_0} + {H_{{\rm{Kondo}}}} + {H_{{\rm{RKKY}}}};\\
H_0=-t\sum\limits_{i\sigma }[ {c_{i\sigma }^ {\dagger}{c_{i+1\sigma }}}+h.c.],\\
H_{{\rm{Kondo}}}=J\sum\limits_{j = 1,2} {{{\bf{S}}_j}\cdot {\bf{s}}\left( {{{\bf{R}}_j}}\right)},\\
H_{{\rm{RKKY}}}=K{{\bf{S}}_1} \cdot {{\bf{S}}_2},
\end{array}
\end{equation}
where $H_0$ describes the non-interacting conduction band with $c_{i\sigma }$ being the annihilation operator of a conduction electron located at the $i$-th site with spin component $\sigma$ and $t$ being the nearest-neighbor hopping integral, $J$ is the short-range (on-site) Kondo coupling between each impurity spin $ {\bf{S}}_j$ and the spin ${\bf s}\left( {{{\bf{R}}_j}} \right)$ of a conduction electron passing by the $j$-th impurity site, and $K$ is the direct exchange interaction between the two impurity spins. Such a TIKM can be realized in nanoscale devices where the observed Kondo signature varies with tunable RKKY interaction\cite{Glazman04,Craig04,Pruser14}. Notice that in realistic materials the direct interaction $K$, which is finite even at vanishing Kondo coupling, can be mediated as a superexchange via other filled orbitals as in some layered $f$-electron compounds\cite{Dai09}.

As shown in Fig. \ref{model}, the two impurities are located $R=|{\bf{R}}_2-{\bf{R}}_1|$ lattice spacings apart. The model exhibits the lattice inversion symmetry with respect to the mirror center of the two impurities. As illustrated by Affleck et al.\cite{Affleck95}, at half filling of the conduction band the model exhibits the first (or second) type of PH symmetry when $R$ is even (or odd). Throughout the present study we fix $t=1/2$ and take half filling of the conduction band, while the periodic boundary condition is imposed and the length of the chain is denoted by $L$. When $K = 0$, the model becomes the one studied by Fye et al. using QMC method \cite{Fye89, Fye94}.

Motivated by the suggestion of Affleck et al.\cite{Affleck95}, we mainly consider the two representative cases, i.e., the interimpurity distance being odd ($R=1,3$) or even ($R=2,4$),
respectively. We note that the case with even $R$ was not thoroughly considered in the previous QMC calculations\cite{Fye89,Fye94}. However, the required first type of PH symmetry is realized only in this situation. We also notice that actually in most of the previous numerical studies \cite{Jones87,Jones88,Fye89,Fye94,Allerdt14} only indirect RKKY interaction is considered, corresponding to $K=0$.

In the present work, the eigenvalues and wave functions of the ground state and low-lying states of Hamiltonian (\ref{eq-hmlt}) were first directly solved by the NORG method for sufficiently large systems. The relevant physical quantities were then calculated. Here the proper physical quantities to characterize the ground state structure of the system and describe its response to external field are respectively the interimpurity spin-spin correlation function $\langle {\bf S}_1\cdot {\bf S}_2 \rangle$ ($\langle\cdots\rangle$ denotes ground state expectation value) and impurity staggered susceptibility $\chi_s $ that is defined as ${\chi _{\bf s}} = \int_0^\infty  {d\tau \left\langle {\left[ {S_1^z\left( \tau  \right) - S_2^z\left( \tau  \right)} \right]\left[ {S_1^z - S_2^z} \right]} \right\rangle }$ with $S_j^z(\tau)=e^{\tau H}S_j^z e^{-\tau H}$. To monitor the change of physical property of the system upon varying couplings $K$ and $J$, we thus calculated $\langle {\bf S}_1\cdot {\bf S}_2\rangle$ and $\chi_s$ as functions of $K$ and $J$. The corresponding quantum phases were achieved by extrapolating $\langle {\bf S}_1\cdot {\bf S}_2\rangle$ and $\chi_s$ to the thermodynamic limit, namely $1/L\rightarrow 0$, as described in Appendix C for details.

\begin{figure}[b]
\includegraphics[width=8.0cm]{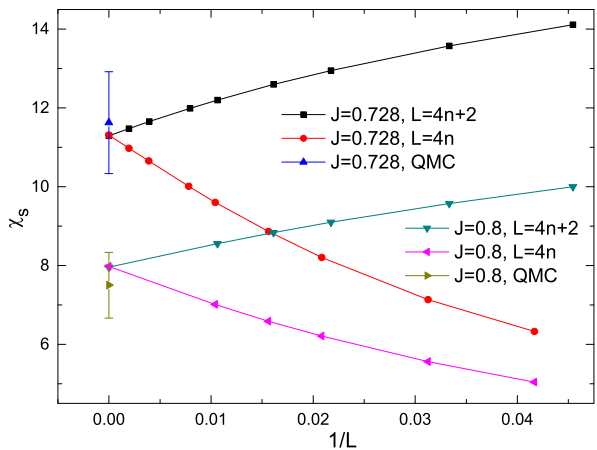}
\caption{(Color online) Finite size extrapolation of impurity staggered susceptibility ${\chi _s}$ at $K = 0$ and at $J = 0.728$ and $0.8$ respectively for $R = 1$. Here $L = 22,~ 24,~ 30,~ ...,~ 510$, and $512,$ respectively. The endpoint at $1/L = 0$ of each curve is determined by a quadratic polynomial fit of the leftmost four points of that curve calculated by NORG. Here the NORG results are consistent well with those of Fye's QMC\cite{Fye94}.}
\label{chi.R1K0}
\end{figure}

As a benchmark test, we carried out the calculations for a set of parameters with $K=0$, $R=1$, $J=0.728$ and $0.8$, and with varying lattice sizes $L$, respectively. Figure \ref{chi.R1K0} shows the calculated impurity staggered susceptibilities along with finite size extrapolations to the thermodynamic limit. As we see, the results in the thermodynamic limit fall within the error bars of those calculated by the early QMC \cite{Fye94}. This also confirms the early QMC results. We remind that the case of $K\neq 0$ was not studied by QMC calculations since this would introduce technical difficulties in QMC. By contrast, there is no extra difficulty by using the NORG approach, as shown below.

\begin{figure}
\includegraphics[width=8.0cm]{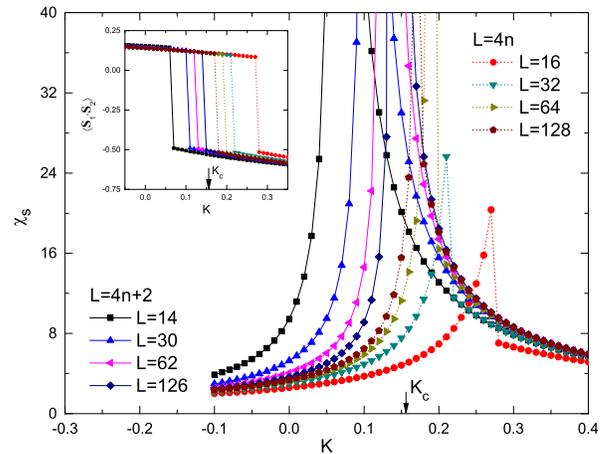}
\caption{(Color online) Impurity staggered susceptibility ${\chi _s}$ and interimpurity spin-spin correlation $\left\langle {{{\bf{S}}_1}\cdot {{\bf{S}}_2}} \right\rangle $ (inset) as functions of $K$ for $J = 1$ and $R = 2$ with different lengths $L$. It is similar to the case of $J = 1$ and $R = 4$ (not shown). For $L = 4n + 2$ represented by solid lines \cite{note2}, ${\chi _s}$ shows a divergent peak at $K = K_c^{{\rm {odd}}}(L)$ which moves to the right and finally converges to $K_c \equiv K_c(L = \infty) = 0.1555$ as $L \rightarrow \infty$\cite{suppl}. And for $L = 4n$ represented by dotted lines, ${\chi _s}$ shows a peak (with discontinuity) at $K = K_c^{{\rm {even}}}(L)$ which moves to the left and finally converges to the same $K_c$ (determined in the $L = 4n + 2$ case) as $L \rightarrow \infty$ and meanwhile the peak height increases and finally diverges. $\left\langle {{{\bf{S}}_1}\cdot {{\bf{S}}_2}} \right\rangle $ jumps where ${\chi _s}$ peaks. The discontinuity of $\left\langle{{{\bf{S}}_1} \cdot {{\bf{S}}_2}} \right\rangle $ indicates a quantum phase transition, which is further characterized as a quantum critical point by the divergence of ${\chi _s}$. } \label{R2J1}
\end{figure}

We first study the cases of interimpurity distance $R=2$ and 4 with Kondo coupling $J=1$ respectively, in which there is the first type of PH symmetry. Figure \ref{R2J1} shows the calculated impurity staggered susceptibility $\chi_s$ and interimpurity spin-spin correlation $\langle{\bf S}_1\cdot{\bf S}_2\rangle$ at zero temperature as functions of $K$ for various lattice sizes. Physically a very large negative $K$, namely a large ferromagnetic coupling, tends to lock the two impurity spins into a triplet, characterized by $\langle{\bf S}_1\cdot{\bf S}_2\rangle$ being positive and close to the triplet value $\frac{1}{4}$. This has been shown in the inset of Fig. \ref{R2J1}. The resulting triplet is then screened by the conduction electrons through the two-stage Kondo effect process\cite{JKW81}. In particular, we see that this tendency extends to $K=0$. This means that the {\it indirect} RKKY interaction, which is due to the second or higher order corrections to the expectation value of $\langle{\bf S}_1\cdot{\bf S}_2\rangle$ between two impurity spins (when $K=0$), is {\it ferromagnetic} at half filling. Thus the ground state for $K=0$ is always in the Kondo regime. In order to reach the border of the interimpurity singlet state regime, we tuned the strength of $K$ in a wider range from ferromagnetic to antiferromagnetic.

\begin{figure}[b]
\includegraphics[width=8.0cm]{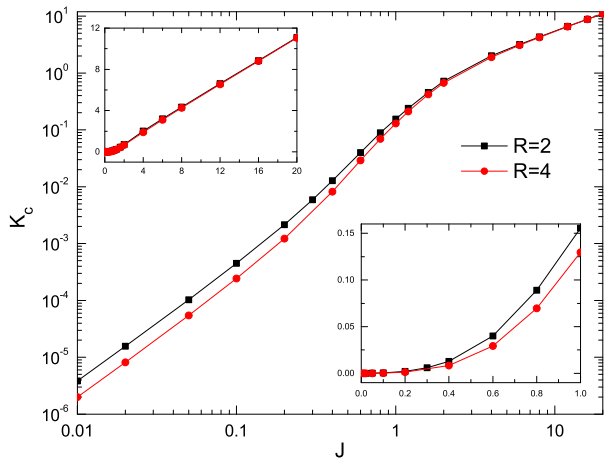}
\caption{(Color online) Phase diagram of direct exchange interaction $K$ versus Kondo coupling $J$: calculated critical point $K_c$ as a function of $J$. This defines a phase boundary separating the Kondo singlet phase (small $K$) from the interimpurity singlet phase (large $K$) at zero temperature. $K_c(J)$ is quadratic at the small $J$ limit (lower inset) and linear at the large $J$ limit (upper inset).} \label{Kc}
\end{figure}

For a large positive $K$, namely a large antiferromagnetic coupling, the ground state shall escape from the Kondo screening phase, entering the interimpurity spin singlet phase. This is clearly seen from the interimpurity spin-spin correlation $\langle{\bf S}_1\cdot{\bf S}_2\rangle $ (the inset of Fig. \ref{R2J1}), which is negative and approaching the singlet value $-\frac{3}{4}$ on the large positive $K$ side. In this case, there is the development of interimpurity singlet driven by the antiferromagnetic correlation. In the intermediate regime of $K\sim 0.15$, however, we find a rather sharp peak in the impurity staggered susceptibility. With increasing lattice size, the peak goes diverging and the position of the peak converges to $K_c$, indicating that the divergence of impurity staggered susceptibility is an intrinsic property free of the finite size effect. Furthermore, when $K$ is around the location of the impurity staggered susceptibility peak, the corresponding interimpurity spin-spin correlation $\langle{\bf S}_1\cdot{\bf S}_2\rangle$ exhibits a concomitant jump across the onset of antiferromagnetic correlation signed by $\langle \mathbf {S}_1\cdot \mathbf {S}_2\rangle$ beginning to be negative. This feature indicates a quantum phase transition taking place at $K=K_c$ from the Kondo singlet regime to the interimpurity singlet regime, which is further characterized as a quantum critical point by the divergence of the impurity staggered susceptibility.

\begin{figure}[t]
\includegraphics[width=8.0cm]{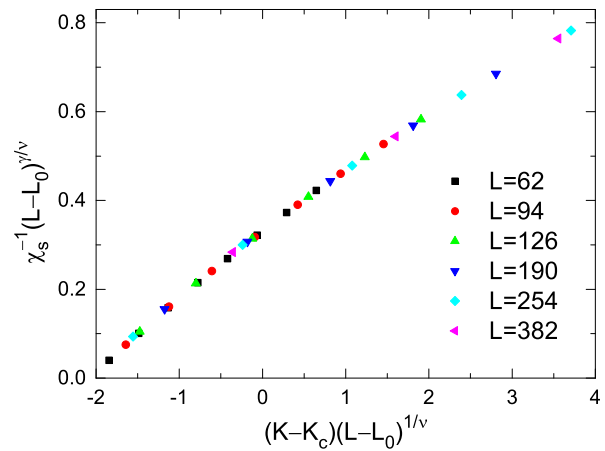}
\caption{(Color online) Finite-size data collapse for the case of $J = 1$ and $R = 2$ with different lengths $L$. This gives rise to $\chi_s \sim |K - K_c|^{-\gamma}$ with $\gamma = 0.60 \pm 0.01$. $L_0 = -9.2$ and $\nu = 1$ are determined by fitting data with formula $|K_c(L) - K_c| \sim (L - L_0)^{-1/\nu}$ \cite{suppl}.} \label{criexp}
\end{figure}

The location of the critical point $K_c$ can be precisely determined by fitting the location of the divergent peak in the large $L$ limit, as illustrated in Appendix C. Figure \ref{Kc} shows the critical point $K_c$ as a function of Kondo coupling $J$, which defines a phase boundary separating the Kondo singlet phase ($K<K_c$) from the interimpurity singlet phase ($K>K_c$). As we see, the phase diagrams for interimpurity distance $R=2$ and $R=4$ are basically the same, while the function $K_c(J)$ shows a quadratic behavior at the small $J$ limit and linear behavior at the large $J$ limit.

To quantify the quantum criticality, we elaborately calculated the impurity staggered susceptibility $\chi_s(K)$ as a function of $K$ in a range centered at the critical point $K_c$ with different system sizes, reported in Fig. \ref{criexp}. Using a standard numerical technique\cite{Sandvik10}, we can derive that $\chi_s(K)$ diverges towards $K_c$ in power law rather than logarithmically in the thermodynamic limit, namely $\chi_s(K)\sim |K-K_c|^{-\gamma}$ with $\gamma=0.60\pm 0.01$. Hence the critical divergence is faster than the logarithmic one but still slower than $|K-K_c|^{-1}$, indicative of a continuous phase transition in computational studies\cite{Sandvik10}. This exponent $\gamma$ has not been reported in literatures to our knowledge.

\begin{figure}
\includegraphics[width=8.0cm]{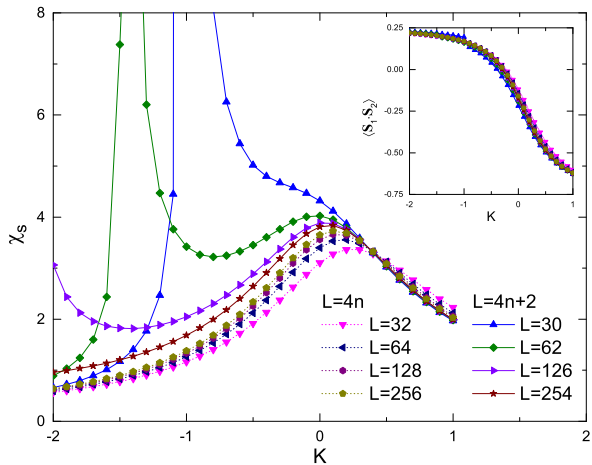}
\caption{(Color online) Impurity staggered susceptibility ${\chi_s}$ and interimpurity spin-spin correlation $\left\langle {{{\bf{S}}_1}\cdot {{\bf{S}}_2}} \right\rangle $ (inset) as functions of $K$ for $R = 1$. $J = 1$ is fixed with different lengths $L$. For $L = 4n + 2$, ${\chi _s}$ looks diverging at some value of $K = K_c$. But the diverging tendency is suppressed and $K_c$ moves to $-\infty$ as $L \rightarrow \infty$ \cite{note2,suppl}. Therefore, the ${\chi _s}$ curves for $L = 4n + 2$ and $L = 4n$ both converge to the same one in the thermodynamic limit $n\rightarrow\infty$. The case of $R = 3$ is similar (not shown). } \label{R1J1}
\end{figure}

As a comparison, we likewise study the cases of interimpurity distance $R=1$ and 3 with Kondo coupling $J=1$ respectively. There is now the second type of PH symmetry, which however does not guarantee any phase transition. In Fig. \ref{R1J1} we report the calculated impurity staggered susceptibility $\chi_s$ and interimpurity spin-spin correlation $\langle \mathbf {S}_1\cdot \mathbf {S}_2\rangle$ as functions of the direct exchange interaction $K$. Similar to the cases of $R=2$ and 4, for a large negative (positive) $K$, the $\langle \mathbf {S}_1\cdot \mathbf {S}_2\rangle$ approaches the triplet (singlet) value, namely $\frac{1}{4}$ ($-\frac{3}{4}$) (the inset of Fig.\ref{R1J1}), corresponding to the Kondo (interimpurity) singlet regime. However, there is no any jump or discontinuity for $\left\langle {{{\bf{S}}_1} \cdot {{\bf{S}}_2}} \right\rangle $ upon $K$ varying. Actually, as shown in the inset of Fig. \ref{R1J1}, the change of $\left\langle {{{\bf{S}}_1} \cdot {{\bf{S}}_2}} \right\rangle $ from the Kondo singlet regime to the interimpurity singlet regime is very smooth across the middle value $-\frac{1}{4}$ between the singlet and triplet values, indicating co-existence of the Kondo and interimpurity singlets in this intermediate regime, in other words, a crossover taking place. This can be further confirmed by examining the behavior of impurity staggered susceptibility $\chi_s$ as a function of $K$ as follows. In addition, the $\left\langle {{{\bf{S}}_1} \cdot {{\bf{S}}_2}} \right\rangle $, being negative at $K=0$, shows that the {\it indirect} RKKY interaction is antiferromagnetic at $K=0$, contrary to the cases of $R=2$ and 4.

From Fig. \ref{R1J1} we see that the ${\chi _s}(K)$ curves for $L = 4n$ converge to a smooth and non-divergent one as $L \rightarrow \infty$ \cite{note2}. In contrast, each ${\chi _s}(K)$ curve for finite $L = 4n + 2$ seemingly has a divergent peak. However, further calculations \cite{suppl} show that the value of $K$ at which ${\chi_s}(K)$ diverges for $L = 4n + 2$ moves to the negative infinity as $L \rightarrow \infty$. It turns out that eventually the ${\chi_s}(K)$ curves for both $L = 4n + 2$ and $L = 4n$ converge to the same smooth and non-divergent one in the thermodynamic limit. This indicates the absence of a quantum critical point. This conclusion is in agreement with the previous QMC calculations for the case of $R = 1$ (or $R = 3$).

\section{Discussion and Outlook}

The studied TIKM offers a new ingredient, namely the interimpurity exchange $K$, which competes with the single-ion Kondo effect of individual impurities. At zero temperature, two different regimes of the ground state, i.e., the Kondo singlet and the inter-impurity singlet, are respectively realized when $K$ is varied from the large ferromagnetic to the large antiferromagnetic. Crossover between these two regimes is expected to be a generic feature since the relevant degrees of freedom due to quantum impurities are finite. Thus it comes as a surprise when an unstable non-Fermi liquid fixed point or a quantum critical point separating the two regimes was firstly evidenced by the NRG studies\cite{Jones87,Jones88}. The existence of such a critical point in the TIKM  has been then debated by various approaches even in the presence of PH symmetry\cite{Fye89,Fye94,Sakai90,Sakai92,Silva96,Allerdt14}.

Using the NORG calculations we have demonstrated that there is indeed a quantum critical point for a TIKM with the first type of PH symmetry. The criticality is characterized by the power-law divergence of the impurity staggered susceptibility $\chi_s$, concomitantly accompanied by a jump of the interimpurity spin-spin correlation $\left\langle {{{\bf{S}}_1} \cdot {{\bf{S}}_2}} \right\rangle$, and thus belongs to a genuine impurity quantum phase transition where the impurity contribution to the ground state energy becomes singular \cite{Vojta06,Bayat14,note3}. Moreover, as presented in above sections, we have also shown that to resolve this longstanding issue depends on not only the required PH symmetry but also the tunable direct RKKY interaction. Meanwhile, the position of critical point is closely related to the lattice details. The newly developed NORG method, which keeps lattice details and is free of decomposition and finite temperatures, makes examination on all these aspects possible.

Here we would like to remind that in the previous NRG studies on TIKM (or the corresponding two-impurity Anderson model) the occurrence of a critical point requires vanishing even/odd channel asymmetry\cite{Silva96} or energy-independence of coupling constants \cite{Sakai90,Sakai92}. However, the even/odd channel asymmetry or energy-dependence of coupling constants is an inevitable consequence of decomposition when mapping a lattice TIKM onto a continuous model in momentum or energy space. On the other hand, the phase shift argument \cite{Jones89b,Affleck95,note1} for a quantum phase transition in a TIKM does not reply on even/odd channel asymmetry or energy-independence of coupling constants. Therefore, our NORG study on a standard lattice TIKM directly confirms the long-cherished prediction of the link between the first type of PH symmetry and the two-impurity Kondo critical point \cite{Affleck95}.

The present study clearly shows that the previous numerical studies actually do not contradict each other for their own sake because the discrepancies are mainly due to the fact that they adopted different parameter regions or approximations. In particular, the energy-dependence of the coupling constants appeared in these studies should not be an obstacle for studying the critical point provided that the first type of PH symmetry and antiferromagnetic interimpurity interaction are realized. As the required symmetry occurs at half filling in a lattice model with reflection symmetry about a site (or interimpurity distance is even), the two-impurity Kondo critical point should be observed experimentally in realistic materials with tunable interimpurity exchange interaction.

Finally, the NORG method developed here can be also used to solve the multi-impurity models in two or three dimensions\cite{note0}, or used as impurity solvers in the dynamical mean-field theory. Therefore we expect that the NORG method can be used to study other challenging problems in correlated electron systems.

\begin{acknowledgments}

This work is supported by National Natural Science Foundation of China (Grant Nos. 91121008 and 11190024) and National Program for Basic Research of MOST of China (Grant No. 2011CBA00112). J.D. was also supported in part by the Qianjiang Scholarship of Zhejiang Province at Hangzhou Normal University (under Grant No. 2012QDL037). Computational resources were provided by the Physical Laboratory of High Performance Computing in RUC and National Supercomputer Center in Guangzhou with Tianhe-2 Supercomputer.

\end{acknowledgments}

\appendix

\section{Method}

For a quantum impurity model, by generalizing quantum renormalization group approach into natural orbitals space through iterative orbital rotations, we can realize a numerical many-body approach, namely the NORG\cite{He14}, with polynomial, no longer exponential, computational complexity ($O(N_{bath}^4)$) in the number of electron bath sites ($N_{bath}$). It turns out that dozens, or even hundreds, of bath sites can be dealt with in practice. In addition, the NORG works in a Hilbert space constructed from a set of natural orbitals, thus it can work on a quantum impurity model with any lattice topological structure. In the present work, we have improved the efficiency of the NORG method. The computational complexity is further reduced from $O(N_{bath}^4)$ to $O(N_{bath}^3)$, in which the impurity orbitals are no longer involved into orbital rotations, and hence the transformed Hamiltonians become simpler. This enables us to calculate the eigenvalues and wave functions of ground states and low-lying excitations for larger system sizes. The largest system size we have reached is $L = 1022$ (see Appendix C). Errors of the NORG are controllable. In this paper the relative errors of all data calculated by the NORG are less than $10^{-3}$.

\section{Phase shift argument and particle-hole symmetry}

We reiterate the phase shift argument for a phase transition between the collective Kondo screening singlet state and the RKKY interaction-locked interimpurity singlet state in the presence of the first type of particle-hole (PH) symmetry\cite{Affleck92,Affleck95}. In the field theory treatment of the two-impurity Kondo model (TIKM), a decomposition of electron field into even/odd components ( $\psi_{e,E},\psi_{o,E}$) is introduced (the spin degree of freedom is implied), both components are dependent on the quasi-particle energy $E$ measured from the Fermi energy $E_F$ which is set to be zero. The obtained effective Hamiltonian, $H=H_0+H_K$, takes a general form as follows,
\begin{eqnarray}
H_0&=&\int dE E \left\{\psi^{\dagger}_{e,E} \psi_{e,E}
+\psi^{\dagger}_{o,E}\psi_{o,E}\right\},\\
H_K&=&\int dE dE' \left\{
g_{1}\psi^{\dagger}_{e,E}\vec{\sigma}\psi_{e,E'}
\cdot ({\bf S}_1+{\bf S}_2)\right\}\\
&+&\int dE dE' \left\{
g_{2}\psi^{\dagger}_{o,E}\vec{\sigma}\psi_{o,E'}
\cdot ({\bf S}_1+{\bf S}_2)\right\}\nonumber \\
&+&\int dE dE' \left\{
g_{3}\psi^{\dagger}_{e,E}\vec{\sigma}\psi_{o,E'}
\cdot ({\bf S}_1-{\bf S}_2)\right\}\nonumber \\
&+&\int dE dE' \left\{
g_{4}\psi^{\dagger}_{o,E}\vec{\sigma}\psi_{e,E'}
\cdot ({\bf S}_1-{\bf S}_2)\right\},\nonumber
\end{eqnarray}
where $g_{i}$ ($i=1,2,3,4$) are energy-dependent coupling constants reflecting the lattice details. The first type of PH transformation of the fields implies
\begin{eqnarray}
\psi_{e,E}\rightarrow\psi^{\dagger}_{e,-E}\nonumber\\
\psi_{o,E}\rightarrow\psi^{\dagger}_{o,-E},
\end{eqnarray}
while the second type of PH symmetry implies
\begin{eqnarray}
\psi_{e,E}\rightarrow\psi^{\dagger}_{o,-E}\nonumber\\
\psi_{o,E}\rightarrow\psi^{\dagger}_{e,-E}.
\end{eqnarray}

\begin{figure}
\includegraphics[width=8.0cm]{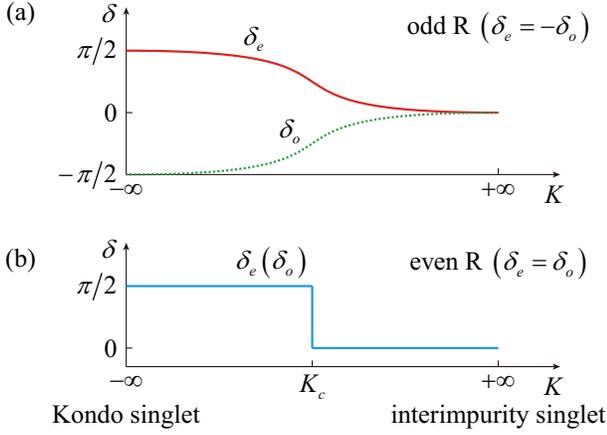}
\caption{(Color online) Schematic view of the phase shift $\delta_{e(o)}$ in the even (odd) channel of the conduction band at the Fermi energy.} \label{shift}
\end{figure}

In the Fermi liquid picture, the zero temperature fixed point can be characterized by the phase shifts ($\delta_e, \delta_o$) for both components at the Fermi energy $E=0$. This amounts to the following relations between the incoming and outgoing operators
\begin{eqnarray}
\psi^{{\rm out}}_{e,E}=e^{2i\delta_e}\psi^{{\rm in}}_{e,E}\nonumber\\
\psi^{{\rm out}}_{o,E}=e^{2i\delta_o}\psi^{{\rm in}}_{o,E}.
\end{eqnarray}
Hence $\delta_e=\delta_o$ in the first type of PH symmetry, and $\delta_e+\delta_o=0$ in the second type of PH symmetry. In the limit of $K\rightarrow -\infty$, both channels are in the Kondo screening phase, indicating $\delta_{e,o}=\pm\pi/2$. While, in the limit of $K\rightarrow \infty$, the Kondo effect is completely suppressed by interimpurity correlation, leading to
$\delta_e=\delta_o=0$. Therefore, by varying $-\infty<K<\infty$, a phase transition must take place at certain value of $K_c$ when the first type of PH is maintained, as sketched in Fig. \ref{shift}(b). By contrast, no phase transition is guaranteed for the second type of PH symmetry (Fig. \ref{shift}(a))

Here we would like to remind that for a lattice model with inversion symmetry and specified Kondo coupling at half filling, the first or second type of PH symmetry will be preserved when the interimpurity separation $R$ is even or odd, respectively. In general, a long-range hybridization or Kondo coupling can be introduced in order to maintain the required PH symmetry. The observation described in the main text indicates that a TIKM with the simplest on-site Kondo coupling which preserves the required PH symmetry is enough to support an emergent quantum critical point.

\section{Complementary numerical results}

\begin{figure}
\includegraphics[width=8.0cm]{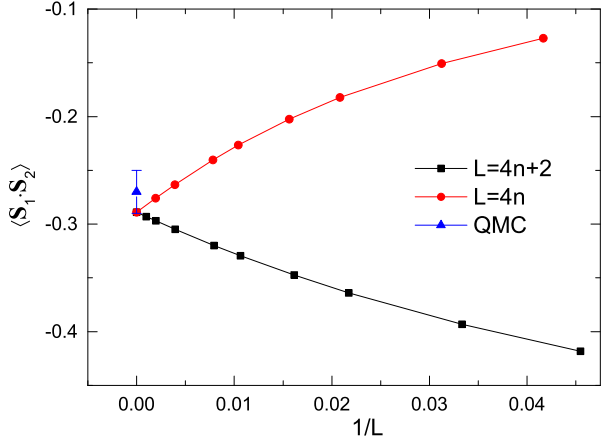}
\caption{(Color online) $\left\langle {{{\bf{S}}_1} \cdot {{\bf{S}}_2}} \right\rangle $ for $R = 1$, $J = 0.728$, and $K = 0$ at zero temperature. $L = 22, 24, 30, ..., 1022$. The endpoint at $1/L = 0$ of each curve is determined by a quadratic polynomial fit of the leftmost four points of that curve calculated by the NORG. Our NORG result is consistent well with that of Fye's QMC\cite{Fye94}.}
\label{S1S2.R1J0.728K0}
\end{figure}

{\it Impurity spin-spin correlation.} Another benchmark test for our NORG method is the impurity spin-spin correlation, $\left\langle {{{\bf{S}}_1} \cdot {{\bf{S}}_2}} \right\rangle $, in the ground state for $R = 1$, where the QMC result is available.  We choose $J = 0.728$ and $K = 0$, the same parameters as given in Fye's QMC study\cite{Fye94}. The calculated result is shown in Fig. \ref{S1S2.R1J0.728K0}. The system size varies as $L = 22, 24, 30, ..., 1022$. The endpoint at $1/L = 0$ of each curve is determined by a quadratic polynomial fit of the leftmost four points of that curve calculated by the NORG. Our NORG result is consistent well with that of Fye's QMC\cite{Fye94}. The relative difference between the two endpoint values is just as small as $1.4 \times 10^{-4}$, which highlights the high accuracy of the NORG method.

\begin{figure}
\includegraphics[width=8.0cm]{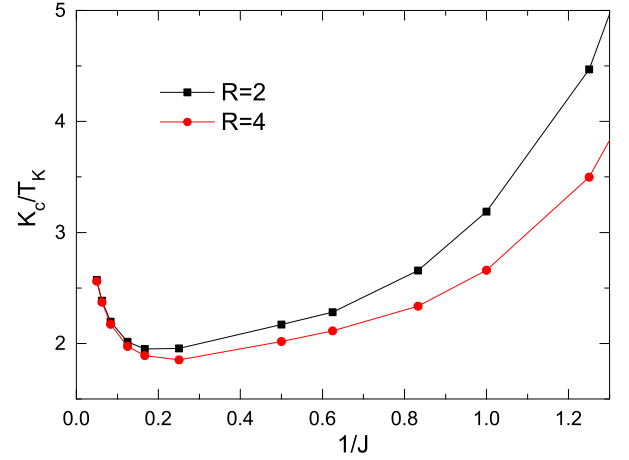}
\caption{(Color online) $K_c / T_K$ versus $1/J$. In the NRG studies\cite{Jones88,Jones89a}, $K_c / T_K$ was predicted to be a constant, about 2.2.} \label{KcTk}
\end{figure}

{\it $K_c/T_K$ when $R=2$ and $4$.} To compare with the early NRG result, we calculate the ratio $K_c/T_K$, where $T_K$ is the single-ion Kondo temperature defined as\cite{Fye94}
\begin{equation}\label{eq-TK}
T_K = D(\rho J)^{1/2} \exp (-1/\rho J),
\end{equation}
where $D$ ($= 2$ here) is the conduction electron bandwidth and $\rho$ ($ = 1 / \pi$ here) the conduction electron density of states at the Fermi level. $K_c/T_K$ is then plotted in Fig.
\ref{KcTk}. While the generic behavior is similar for $R=2$ and $R=4$, quantitative difference between them appears when $J<5$. It is seen that in a wide range of $J$, $K_c / T_K$ is close to the value predicted by the NRG studies, $K_c / T_K \approx 2.2$.\cite{Jones88,Jones89a}

\begin{figure}
\includegraphics[width=8.0cm]{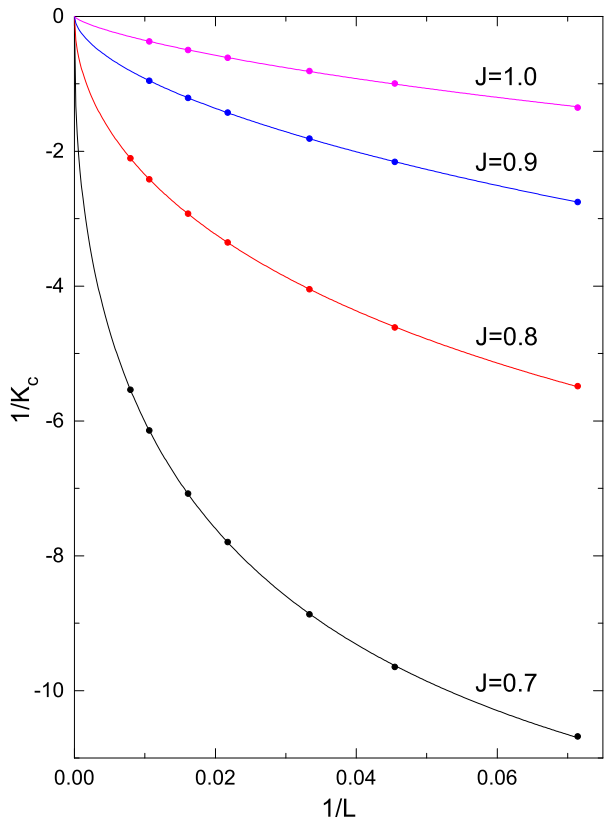}
\caption{(Color online) The value of $K = K_c$ at which ${\chi _s}$ diverges for $R = 3$ and $L = 4n + 2$. Dots denote the results calculated by the NORG. Lines are determined by fitting the dots with the formula $K_c = -a (L-c)^b$, where $a$, $b$, and $c$ are fitting parameters. The $K_c$ behavior for $R = 1$ is similar (not shown). This figure clearly shows that $K_c$ decreases to $-\infty$ as $L \rightarrow \infty$.}\label{Kc.R3}
\end{figure}

\begin{figure}
\includegraphics[width=8.0cm]{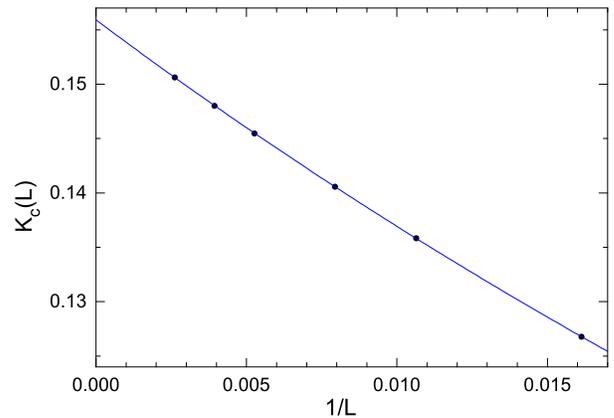}
\caption{(Color online) The value of $K = K_c(L)$ at which ${\chi _s}$ diverges for $R = 2$, $J = 1$, and $L = 4n + 2$. The dots are calculated by NORG. The line is determined by fitting the dots with formula $K_c(L) = K_c - a(L - L_0)^{-1}$, where $K_c$, $a$, and $L_0$ are fitting parameters. This perfect fitting shows that $|K_c(L) - K_c| \sim (L - L_0)^{-1/\nu}$ with $\nu = 1$.}\label{KcL}
\end{figure}

\begin{figure*}
\includegraphics[width=12.cm]{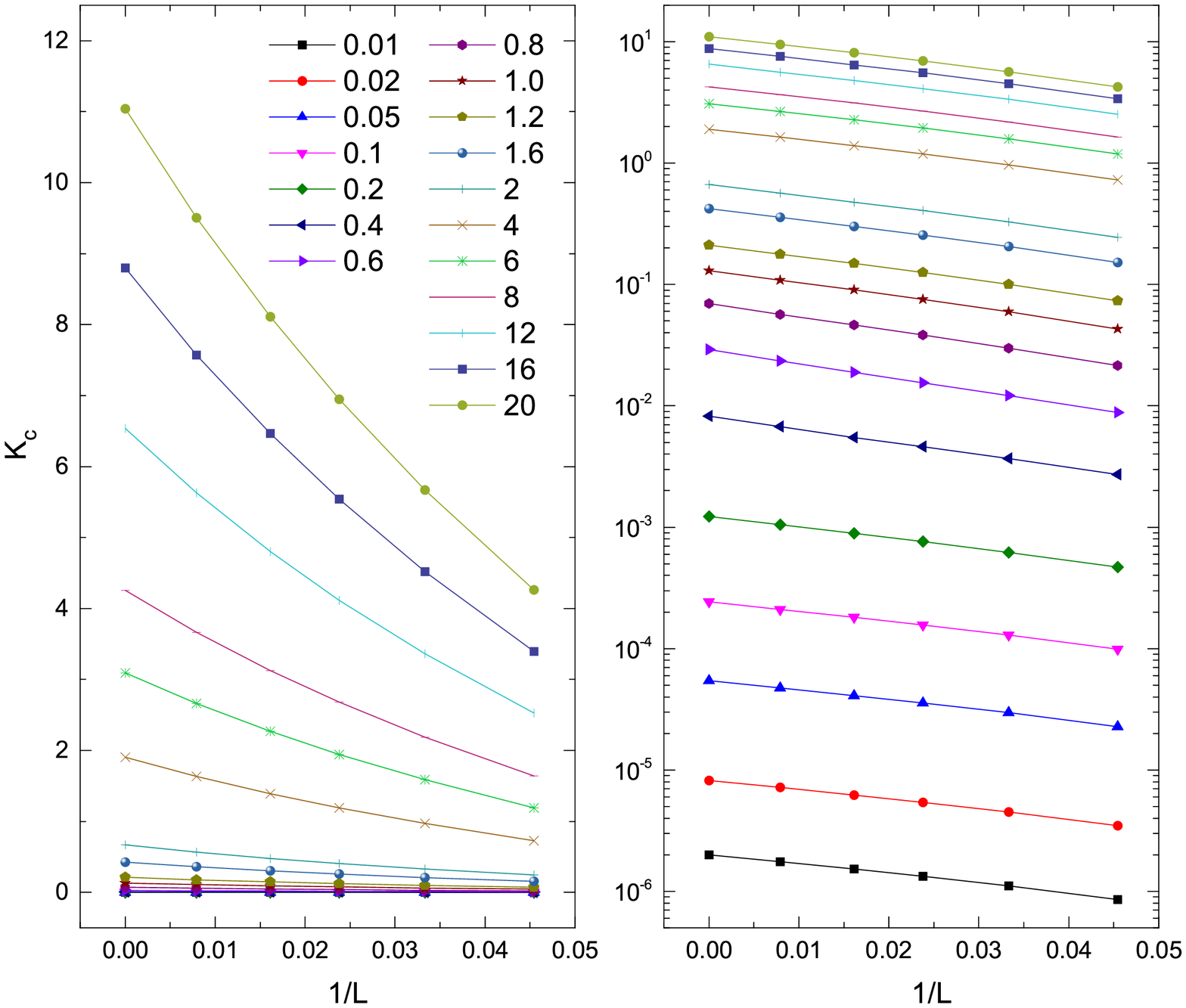}
\caption{(Color online) Finite size extrapolation of $K_c$ for $R = 4$ and various values of $J$. $L = 22, 30, 42, 62$, and $126$. The case for $R = 2$ is similar (not shown). The endpoint at $1/L = 0$ of each curve is determined by fitting the NORG data with formula $K_c(L) = K_c - a(L - L_0)^{-1}$, where $K_c$, $a$, and $L_0$ are fitting parameters. The fitting quality is very well. Figure \ref{KcL} is a typical example.} \label{KcR4J1}
\end{figure*}

{\it Determination of $K_c$ when $R$ is even}. For even R the TIKM has a quantum critical point at $K = K_c$ where the impurity staggered susceptibility ${\chi _s}$ diverges. For finite system size $L = 4n + 2$, ${\chi _s}$ diverges at $K = K_c(L)$. We extract these $K_c(L)$ from ${\chi _s}$ curves calculated by NORG for different values of $L$ and extrapolate it to $L = \infty$ by fitting. To present the fitting details we show the typical case of $J = 1$ in Fig. \ref{KcL}. The perfect fitting shows that $|K_c(L) - K_c| \sim (L - L_0)^{-1/\nu}$ with $\nu = 1$. For a series of values of $J$, we show the result in Fig. \ref{KcR4J1}. While $J$ spans across three orders of magnitude, $K_c$ across seven orders of magnitude. This shows the robustness of the NORG method.

{\it Suppression of susceptibility peak when $R$ is odd.} The impurity staggered susceptibility $\chi_s$ shows a sharp peak in the case of $R=3$ (or $R=1$) when $L=4n+2$. This feature seems different to what happens when $L=4n$ where no sharp peak is detected. In order to clarify whether this is an intrinsic property of the ground state in the thermodynamic limit $L\rightarrow \infty$, we plot the peak position $K_c$ by varying the lattice size $L$ for different values of $J$ (Fig. \ref{Kc.R3}). The result suggests that in the limit $L\rightarrow\infty$, the peak position $K_c$ of ${\chi _s}$ for $L = 4n + 2$ moves to negative infinity and ${\chi _s}$'s for both $L = 4n + 2$ and $L = 4n$ converge to the same one with a non-divergent behavior in the thermodynamic limit. Therefore, no phase transition takes place in the case of $R=3$ (or $R=1$).


\begin{thebibliography}{99}
\bibitem{note1}
In some variants of the TIKM the conduction electron channels are
doubled such as in double quantum dots while the charge transfer
between dots is prohibited\cite{Zarand06}.
\bibitem{suppl}
A brief illustration of the phase shift argument and some complementary numerical results are given in Appendix B and C.
\bibitem{note0}
Here we follow nothing but the convention to choose a one-dimensional lattice version of the TIKM system as a representative mode system to study\cite{Fye89,Fye94}. As presented in Ref. \onlinecite{He14}, the effectiveness of the NORG is essentially irrespective of a model's topological structure. In other words, the NORG can handle one-dimensional, two-dimensional, and three-dimensional systems with any geometrical structures in the same efficient way. Higher dimensional systems will be studied subsequently.
\bibitem{note2}
For one dimensional tight binding model $H_0$ in Eq. (\ref{eq-hmlt}) with periodic boundary condition and with $2n$ sites ($n$ being an integer), its single-particle energy spectrum consists of a non-degenerate lowest energy level, a non-degenerate highest energy level, and in-between $n-1$ doubly degenerate energy levels. Thus, at half filling the system is of open shell (namely metallic-like behavior) if $n$ is even, while it is of closed shell (namely semiconducting-like behavior) if $n$ is odd. It turns out that the model described by Eq. (\ref{eq-hmlt}) respectively with $4n$ and $4n+2$ sites shows a much different behavior when $n$ is small (see Fig. 5). Nevertheless, the present calculations show that when $n\rightarrow \infty$ in both cases the model realizes the same thermodynamic limit.
\bibitem{note3}
Quantum impurity models involve a finite number of local impurity degrees of freedom and infinite number of bulk degrees of freedom. The coupling between them permits sharp transitions at $T=0$ and the impurity contribution to the free energy is singular at the critical point \cite{Vojta06}. It is also proposed that the Schmidt gap of the entanglement spectra of impurity systems can serve as an alternative order parameter as recently manifested in a dimerization spin chain with two impurity spins\cite{Bayat14}. In our present study, this feature is clearly manifested in the jump of the interimpurity spin-spin correlation. We can envision that by increasing the number of impurities to infinity the jump could be replaced by a continuous transition as in the Kondo lattice case with a local quantum critical point.\cite{Si01}
\end{thebibliography}
\end{document}